# The magic nature of $^{132}$Sn explored through the single-particle states of $^{133}$Sn

K. L. Jones[1,2], A. S. Adekola[3], D. W. Bardayan[4], J. C. Blackmon[4], K. Y. Chae[1], K. A. Chipps[5], J. A. Cizewski[2], L. Erikson[5], C. Harlin[6], R. Hatarik[2], R. Kapler[1], R. L. Kozub[7], J. F. Liang[4], R. Livesay[5], Z. Ma[1], B. H. Moazen[1], C. D. Nesaraja[4], F. M. Nunes[8], S. D. Pain[2], N. P. Patterson[6], D. Shapira[4], J. F. Shriner Jr[7], M. S. Smith[4], T. P. Swan[2,6] & J. S. Thomas[6]

[1]Department of Physics and Astronomy, University of Tennessee, Knoxville, Tennessee 37996, USA. [2]Department of Physics and Astronomy, Rutgers University, New Brunswick, New Jersey 08903, USA. [3]Department of Physics and Astronomy, Ohio University, Athens, Ohio 45701, USA. [4]Physics Division, Oak Ridge National Laboratory, Oak Ridge, Tennessee 37831, USA. [5]Physics Department, Colorado School of Mines, Golden, Colorado 80401, USA. [6]Department of Physics, University of Surrey, Guildford, Surrey GU2 7XH, UK. [7]Department of Physics, Tennessee Technological University, Cookeville, Tennessee 38505, USA. [8]National Superconducting Cyclotron Laboratory and Department of Physics and Astronomy, Michigan State University, East Lansing, Michigan 48824, USA.

**Atomic nuclei have a shell structure[1] where nuclei with 'magic numbers' of neutrons and protons are analogous to the noble gases in atomic physics. Only ten nuclei with the standard magic numbers of both neutrons and protons have so far been observed. The nuclear shell model is founded on the precept that neutrons and protons can move as independent particles in orbitals with discrete quantum numbers, subject to a mean field generated by all the other nucleons. Knowledge of the properties of single-particle states outside nuclear shell closures in exotic nuclei is important[2–5] for a fundamental understanding of nuclear structure and nucleosynthesis (for example the r-process, which is responsible for the production of about half of the heavy elements). However, as a result of their short lifetimes, there is a paucity of knowledge about the nature of single-particle states outside exotic doubly magic nuclei. Here we measure the single-particle character of the levels in $^{133}$Sn that lie outside the double shell closure present at the short-lived nucleus $^{132}$Sn. We use an inverse kinematics technique that involves the transfer of a single nucleon to the nucleus. The purity of the measured single-particle states clearly illustrates the magic nature of $^{132}$Sn.**

The nuclear shell model[1] explains why particular numbers of protons and/or neutrons (2, 8, 20, 28, 50 and 82, as well as 126 for neutrons) result in additional binding compared with the neighbouring isotopes. Nuclei with these standard magic numbers have comparatively high energies for their first excited $2^+$ state and high energies needed to remove one or two nucleons. Conversely, nuclei just beyond shell closures have low nucleon separation energies. These nucleon separation energies are analogous to ionization energies in atoms: noble gases, with closed shells of electrons, show large discontinuities in their ionization energies compared with neighbouring elements, whereas the alkali metals, having a single electron outside the closed shell, are good electron donors. In the extreme limit of the nuclear shell model the properties, such as the spin and parity, of an odd-mass nucleus are determined solely by the single unpaired nucleon. This assertion is valid to a high degree, and especially so at the magic numbers that correspond to significant gaps in the spacings of the single-particle energies.

Characterizing the nature of single-particle states just outside a double-shell closure is essential in calibrating theoretical models of the nucleus and predicting the properties of the thousands of currently unmeasured nuclei, such as those involved in the astrophysical rapid neutron-capture process, commonly called the r-process[6]. This process is sited in an extremely neutron-rich, high-temperature environment, such as a supernova or a merger of neutron stars. The r-process is responsible for the production of more than half of the elements heavier than iron by means of successive neutron captures on unstable neutron-rich nuclei. The inputs from nuclear structure, such as masses and lifetimes, for r-process simulations come from the known properties of accessible nuclei and nuclear model predictions.

In studying the magic nature of $^{132}$Sn, its properties are compared with those of the nucleus that represents the benchmark of magic nuclei, namely $^{208}$Pb (ref. 7). Some comparisons between $^{208}$Pb and $^{132}$Sn are shown in Fig. 1, including the large energies of the first $2^+$ states in comparison with those of the neighbouring isotopes. The shell closure reveals itself as a large discontinuity, for instance at $^{132}$Sn, where $E_{2+}$ = 4041.2(15) keV is significantly higher than that of the other tin isotopes (about 1200 keV) and drastically larger than that for nearby isotopes of cadmium or tellurium (about 500 keV) (Evaluated Nuclear Structure Data File (ENSDF) database; http://www.nndc.bnl.gov/ensdf/). However, these excitations alone do not prove that a nucleus is magic, because they may reflect other properties such as changes in pairing strength[8]. Another sign of magic nature comes from the sudden decrease in two-neutron separation energies—$S_{2n}$ is shown in Fig. 1b—for the isotopes just beyond the shell closure.

A critical test of the shell closure is to study the single-particle states outside the closed shell. An important metric is the spectroscopic factors ($S$) of single-particle states in the nuclei with one neutron or one proton beyond the double-shell closure. For a good magic nucleus $A$, the single-particle strength for a specific orbital in the $A + 1$ nucleus should be concentrated in one state, resulting in high spectroscopic factors, as opposed to being fragmented through the spectrum of the nucleus.

Situated at the beginning of the neutron 82–126 shell, the single-particle orbitals in $^{133}$Sn are expected to be $2f_{7/2}$, $3p_{3/2}$, $1h_{9/2}$, $3p_{1/2}$, $2f_{5/2}$ and $1i_{13/2}$ (the five bound states are shown in Fig. 1d). Candidates for four of these states have been observed[9,10], with the notable exception of the $p_{1/2}$ and the $i_{13/2}$ orbitals. The experimental values of the excitation energies of single-particle states just outside a shell closure are important benchmarks for shell-model calculations for more exotic nuclei. Experimental investigations of the single-particle nature of $^{133}$Sn have been confined to β-decay measurements[9] and the spectroscopy of prompt γ-rays after the fission of $^{248}$Cf (ref. 10). In this region of the nuclear chart, β-decay preferentially populates high-spin states in the daughter nucleus. In fission fragment spectroscopy both the production of the daughter nucleus of interest and the techniques used to extract information from the plethora of photons emitted from a fission source favour high-spin states. Therefore, none of the previous measurements of $^{133}$Sn were well suited to the study of low-spin states, and none was a direct probe of the single-particle character of the excitations.

One very sensitive technique for studying low angular momentum, single-particle states is by means of a reaction in which a single nucleon is ‘transferred’ from one nucleus to another. These transfer reactions traditionally require a light ion beam striking a target of higher mass. For nuclei far from stability this is not possible, because the target would not live long enough to perform the measurement. Recently these reactions have been performed in inverse kinematics with light-$A$ targets, in particular deuterons in deuterated polyethylene ($CD_2$) targets, and radioactive ion beams[11,12]. These measurements include the pioneering experiment on the long-lived doubly magic nucleus $^{56}$Ni (ref. 13). In a (d,p) reaction in inverse kinematics, a neutron is removed from a deuteron (d) in the target, and is transferred to a beam particle, ejecting a proton (p) that can be detected (see Fig. 2 top left inset). This reaction is ideally suited to the study of low-lying single-neutron states in the final nucleus.

To perform the $^{132}$Sn(d,p) reaction in inverse kinematics, a beam[14] of the short-lived isotope $^{132}$Sn ($t_{1/2}$ = 39.7 s ) was produced at the Holifield Radioactive Ion Beam Facility at Oak Ridge National Laboratory, using the isotope separation online technique. Protons from the Oak Ridge Isochronous Cyclotron bombarded a pressed powder target of uranium carbide, inducing fission. Negative ions of tin were injected into and accelerated by the 25-MV tandem electrostatic accelerator to 630 MeV. The resulting essentially pure (more than 90%) $^{132}$Sn beam bombarded a $CD_2$ reaction target with an effective areal density of 160 μg cm$^{-2}$. Protons emerging from the (d,p) reaction were measured in position-sensitive silicon Oak Ridge Rutgers University Barrel Array (ORRUBA)[15] detectors covering polar angles between 69° and 107° in the laboratory frame. At forward angles, telescopes of ORRUBA detectors consisting of 65-μm or 140-μm $\Delta E$ (energy loss) detectors backed by 1000-μm $E$ (residual energy) detectors were employed to stop elastically scattered $^{12}$C recoils coming from the composite $CD_2$ target, and to allow particle identification. Backwards of the elastic scattering region ($\theta_{lab}$ > 90°), single-layer 1000 μm ORRUBA detectors were used. A microchannel plate detector[16] located downstream of the target chamber provided a timing signal for beam-like recoil particles. The elastic scattering of deuterons from the target was used in the normalization of the transfer reaction cross-sections. These data, taken at forward angles ($\theta_{CM}$ = 28–43°), were dominated by Rutherford scattering, which can be easily calculated. Small corrections (about 6% or less) due to nuclear scattering were included in the analysis of the elastic scattering data. In this way uncertainties in the number of target deuterons and beam ions were greatly decreased in the normalization.

Figure 2 shows the reaction $Q$-value spectrum for the $^{132}$Sn(d,p) reaction as measured at 54° in the centre-of-mass frame. Four clear peaks can be seen, corresponding to the ground state, the known $E_x$ = 854 keV and $E_x$ = 2005 keV excited states, and a previously unobserved state at $E_x$ = 1363 ± 31 keV. The tentative spin-parity assignments for the known states are 7/2$^-$ (presumably $2f_{7/2}$), 3/2$^-$ (presumably $3p_{3/2}$) and $5/2^{-}$ (presumably $2f_{5/2}$), respectively. The initial supposition for the nature of the new state is that it is the hitherto unobserved $3p_{1/2}$ state.

Angular distributions of the protons from single-neutron transfer experiments reflect the orbital angular momentum, *l*, of the transferred nucleon. Because the (d,p) reaction preferentially populates low-*l* single-neutron states, only *p*-wave and *f*-wave states in the region above $^{132}$Sn are expected to be significantly populated in the $^{132}$Sn(d,p) reaction. Angular distributions for the four states measured were extracted from the *Q*-value spectra at different angles by using a four-Gaussian fit. The widths of the peaks were allowed to increase for the higher excited states, reflecting the diminished *Q*-value resolution for low-energy protons. For each state, transfer angular distributions to an $l = 1$ and an $l = 3$ state were calculated in the distorted-wave Born approximation (DWBA) framework, with the use of the code FRESCO[17]. The Reid interaction[18] was used for the deuteron and the finite-range DWBA calculation included full complex remnant in the transfer operator. The optical model potentials were taken from ref. 19, and standard Woods–Saxon parameters for the radius parameter $r = 1.25$ fm (where the radius $R$ is given by $R = rA^{1/3}$) and diffuseness $a = 0.65$ fm for the final bound state were used. Spectroscopic factors were extracted by scaling the DWBA calculation to the data. Figure 3a, b shows the angular distributions for the two lowest states in $^{133}$Sn with the DWBA calculations scaled to reproduce the data. The ground-state data (Fig. 3a) prefer the previously inferred $2f_{7/2}$ assignment to an alternative $3p_{3/2}$ single-particle state available close by. Conversely, the 854-keV state (Fig. 3b) can be distinguished as an $l = 1$ transfer, as would be expected for population of the $3p_{3/2}$ orbital, as opposed to a $2f_{7/2}$ *f*-wave state. Detection of protons with position-sensitive detectors relies on being able to detect a significant signal from both ends of the resistive strip. As the proton energy decreases with increasing excitation energy of the residual $^{133}$Sn nucleus, the area of the strip that is sensitive is reduced. This leads to a decreased angular coverage and angular resolution for the newly observed 1363-keV and 2005-keV states and precludes the determination of significant angular distributions. Spectroscopic factors could nevertheless be determined by comparison of the angle-integrated cross-sections with those from the DWBA calculations. In Fig. 3c and Fig. 3d, respectively, are shown the measured cross-sections over the range of angles for which the 1363-keV and 2005-keV states can be observed, with normalized

DWBA calculations assuming that the state is either the $3p_{1/2}$ or the $2f_{5/2}$. The previously observed ($5/2^-$) state at 2005 keV, shown in Fig. 3d, is expected to be $2f_{5/2}$. The newly observed state with $E_x$ = 1363 keV, shown in Fig. 3c, is a strong candidate for the $3p_{1/2}$ single-particle state that has not previously been observed in $^{133}$Sn and is predicted by the shell model. The $2f_{5/2}$ assignment is unlikely because such a $5/2^-$ state should previously have been observed in the β-decay[9] and fission spectroscopy[10] experiments. Calculations for the 1363-keV state with higher *l*-transfers resulted in cross-sections that were much lower than the data and are thus ruled out on a sum-rule basis; that is, $S \gg 1$ would be required.

Spectroscopic factors are generally sensitive to the choice of the optical potentials (see Supplementary Information). The Strömich potentials[19] provided good fits to the data in terms of shape of the angular distributions and were extracted in a rigorous manner by using elastic scattering and (d,p) data on the last stable tin isotope, $^{124}$Sn. Table 1 summarizes the information on each of the four single-particle states measured here. The *l* values of the lowest two states are well constrained. The asymptotic normalization coefficient (ANC)[20], which is a measure of the normalization of the tail of the overlap function to a Hankel function, is largely independent of the bound-state parameters and is also extracted in this work. The experimental uncertainties in the values of *S* and ANC come from statistics (4–7%), fitting and normalization of the data (10%). An uncertainty of 15% originates from the optical model potentials used in the reaction theory.

Spectroscopic factors can be extracted by using different experimental probes, including electron-induced proton knockout (e,e′p), nuclear-induced nucleon knockout (knockout) and various transfer reactions. There has been some controversy during the past decade regarding probe-dependent discrepancies in *S*. Those extracted from (e,e′p) reactions, for example, consistently have values about 50–60% of the predicted value from shell model calculations[21]. This decrease has been explained as being due to short-range, high-momentum correlations. An *S* extracted from transfer by using standard bound-state parameters should be considered as a relative value. When a radius for the bound state is available, either from a reliable density functional theory valid in this region of the nuclear chart (only ground states) or from experiment (only ground states and for a limited number of nuclei), then an absolute *S* can be extracted. Here we used standard radius and diffuseness parameters; the *S* extracted is therefore relative even though absolute cross-sections were measured. This relative *S* is useful when comparisons are made between values extracted from similar experiments on different isotopes in a careful and consistent manner.

All of the measured states in $^{133}$Sn have large spectroscopic factors $S \approx 1$, and if these were absolute values they would indicate little fragmentation of the single-particle strength. Because they are relative values, the spectroscopic factors for $^{133}$Sn were compared with those obtained for $^{209}$Pb, the core of which is well known for its magic nature. So that we might make a meaningful comparison, finite-range DWBA calculations, similar to those made for the $^{132}$Sn(d,p)$^{133}$Sn reaction, were made for $^{208}$Pb(d,p)$^{209}$Pb using the data from ref. 22 (see Supplementary Information), and were compared with those from ref. 23. Both $^{133}$Sn and $^{209}$Pb have high values of *S* for the lowest-lying states, indicating little fragmentation of the single-neutron strength in these nuclei. In fact, the spectroscopic factors for states above 1 MeV in $^{133}$Sn are consistently larger than their counterparts in $^{209}$Pb, clearly signalling the magic nature of $^{132}$Sn. The resulting spectroscopic factors are shown in Fig. 1c,d.

Here we have determined the purity of the low-spin single-neutron excitations in $^{133}$Sn, namely the $2f_{7/2}$, $3p_{3/2}$, $3p_{1/2}$ and $2f_{5/2}$ orbitals. In addition, the proposed $3p_{1/2}$ state has been measured here and the previously proposed spins of the lowest two states have been confirmed. New calculations of the $^{208}$Pb(d,p)$^{209}$Pb reaction have been made in a manner consistent to those for $^{132}$Sn(d,p)$^{133}$Sn, thus allowing meaningful comparisons to be drawn. The simplicity of $^{132}$Sn, and the single-neutron excitations in $^{133}$Sn, provides a new touchstone needed for extrapolations to nuclei farther from stability, in particular those responsible for the synthesis of the heaviest elements.

1. Mayer, M. G. & Jensen, J. H. D. *Theory of Nuclear Shell Structure* (Wiley, 1955).

2. Barbieri, C. & Hjorth-Jensen, M. Quasiparticle and quasihole states of nuclei around $^{56}$Ni. *Phys. Rev. C* **79,** 064313 (2009).

3. Kartamyshev, M. P., Engeland, T., Hjorth-Jensen, M. & Osnes, E. Effective interactions and shell model studies of heavy tin isotopes. *Phys. Rev. C* **76,** 024313 (2007).

4. Sarkar, S. & Sarkar, M. S. Shell model study of neutron-rich nuclei near $^{132}$Sn. *Phys. Rev. C* **64,** 014312 (2001).

5. Grawe, H., Langanke, K. & Martínez-Pinedo, G. Nuclear structure and astrophysics. *Rep. Prog. Phys.* **70,** 1525–1582 (2007).

6. Cowan, J. J., Thielemann, F.-K. & Truran, J. W. The r-process and nucleochronology. *Phys. Rep.* **208,** 267–394 (1991).

7. Coraggio, L., Covello, A., Gargano, A. & Itaco, N. Similarity of nuclear structure in the $^{132}$Sn and $^{208}$Pb regions: proton–neutron multiplets. *Phys. Rev. C* **80,** 021305(R) (2009).

8. Terasaki, J., Engel, J., Nazarewicz, W. & Stoitsov, M. Anomalous behavior of $2^+$ excitations around $^{132}$Sn. *Phys. Rev. C* **66,** 054313 (2002).

9. Hoff, P. *et al.* Single-neutron states in $^{133}$Sn. *Phys. Rev. Lett.* **77,** 1020–1023 (1996).

10. Urban, W. *et al.* Neutron single-particle energies in the $^{132}$Sn region. *Eur. Phys. J. A* **5,** 239–241 (1999).

11. Kozub, R. L. *et al.* Neutron single particle strengths from the (d,p) reaction on $^{18}$F. *Phys. Rev. C* **73,** 044307 (2006).

12. Thomas, J. S. *et al.* Single-neutron excitations in neutron-rich $^{83}$Ge and $^{85}$Se. *Phys. Rev. C* **76,** 044302 (2007).

13. Rehm, K. E. *et al.* Study of the $^{56}$Ni(d,p)$^{57}$Ni reaction and the astrophysical $^{56}$Ni(p,γ)$^{57}$Cu reaction rate. *Phys. Rev. Lett.* **80,** 676–679 (1998).

14. Stracener, D. W. Status of radioactive ion beams at the HRIBF. *Nucl. Instrum. Methods A* **521,** 126–135 (2004).

15. Pain, S. D. *et al.* Development of a high solid-angle silicon detector array for measurement of transfer reactions in inverse kinematics. *Nucl. Instrum. Methods B* **261,** 1122–1125 (2007).

16. Wiza, J. L. Microchannel plate detectors. *Nucl. Instrum. Methods* **162,** 587–601 (1979).

17. Thompson, I. J. Coupled reaction channels calculations in nuclear physics. *Comput. Phys. Rep.* **7,** 167–211 (1988).

18. Reid, R. V. Local phenomenological nucleon–nucleon potentials. *Ann. Phys.* **50,** 411–448 (1968).

19. Strömich, A. *et al.* (d,p) reactions on $^{124}$Sn, $^{130}$Te, $^{138}$Ba, $^{140}$Ce, $^{142}$Nd, and $^{208}$Pb below and near the Coulomb barrier. *Phys. Rev. C* **16,** 2193–2207 (1977).

20. Pang, D. Y., Nunes, F. M. & Mukhamedzhanov, A. M. Are spectroscopic factors from transfer reactions consistent with asymptotic normalization coefficients? *Phys. Rev. C* **75,** 024601 (2007).

21. Kramer, G. J. & Blok, H. P. & Lapikás, L. A consistent analysis of (e,e′p) and (d,$^3$He) experiments. *Nucl. Phys. A* **679,** 267–286 (2001).

22. Ellegaard, C., Kantele, J. & Vedelsby, P. Particle–vibration coupling in $^{209}$Pb. *Nucl. Phys. A* **129,** 113–128 (1969).

23. Hirota, K., Aoki, Y., Okumura, N. & Tagishi, Y. Deuteron elastic scattering and (d,p) reactions on $^{208}$Pb at $E_d$ = 22 MeV and *j*-dependence of $T_{20}$ in (d,p) reaction. *Nucl. Phys. A* **628,** 547–579 (1998).

**Supplementary Information** is linked to the online version of the paper at www.nature.com/nature.

**Acknowledgements** This research was supported in part by the U.S. Department of Energy under contracts DE-FG02-96ER40983(UTK), DEFG02-96ER40955 (TTU), DE-FG03-93ER40789 (CSM), and DE-AC02-06CH11357 (MSU), the National Nuclear Security Administration under the Stewardship Science Academic Alliances program through DOE Cooperative Agreement DE-FG52-08NA28552 (Rutgers, ORAU, MSU), the National Science Foundation under contracts NSF-PHY-0354870 (Rutgers) and NSF-PHY-0555893 (MSU) and the UK Science and Technology Funding Council under contract number PP/F0000715/1. Oak Ridge National Laboratory is managed by UT-Battelle, LLC, for the U.S. Department of Energy (DOE) under contract DE-AC05-00OR22725.

**Author Contributions** K.L.J., D.W.B., J.C.B., J.A.C., R.L.K., J.F.L., C.D.N., S.D.P., D.S., M.S.S. and J.S.T. designed the experiment and developed the experimental tools and techniques. K.L.J., D.W.B., J.C.B., K.Y.C., R.H., R.L.K., J.F.L., B.H.M., S.D.P. and D.S. set up the experimental equipment, including new, unique detectors and associated electronics. K.L.J., D.W.B., J.C.B., K.Y.C., R.L.K., B.H.M., S.D.P., T.P.S. and J.S.T. developed online and offline analysis software routines and algorithms. K.L.J., A.S.A., D.W.B., J.C.B., K.Y.C., K.A.C., L.E., C.H., R.H., R.K., R.L.K., J.F.L., R.L., Z.M., B.H.M., C.D.N., S.D.P., N.P.P., D.S., J.F.S., M.S.S., T.P.S. and J.S.T. while running the experiment, assessed the quality and performed preliminary analyses of online data. K.L.J., K.Y.C., R.K., R.L.K., B.H.M., S.D.P. and T.P.S. analysed the data and calibrations. K.L.J., D.W.B., J.A.C., R.L.K., F.M.N. and S.D.P. interpreted the data, including theoretical calculations. K.L.J., J.A.C. and F.M.N. wrote the manuscript. K.L.J., D.W.B., J.C.B, K.A.C., J.A.C., R.L.K., J.F.L., F.M.N., S.D.P., J.F.S., M.S.S. and J.S.T. revised the manuscript.

**Author Information** Reprints and permissions information is available at www.nature.com/reprints. The authors declare no competing financial interests. Readers are welcome to comment on the online version of this article at www.nature.com/nature. Correspondence and requests for materials should be addressed to K.L.J. (kgrzywac@utk.edu).

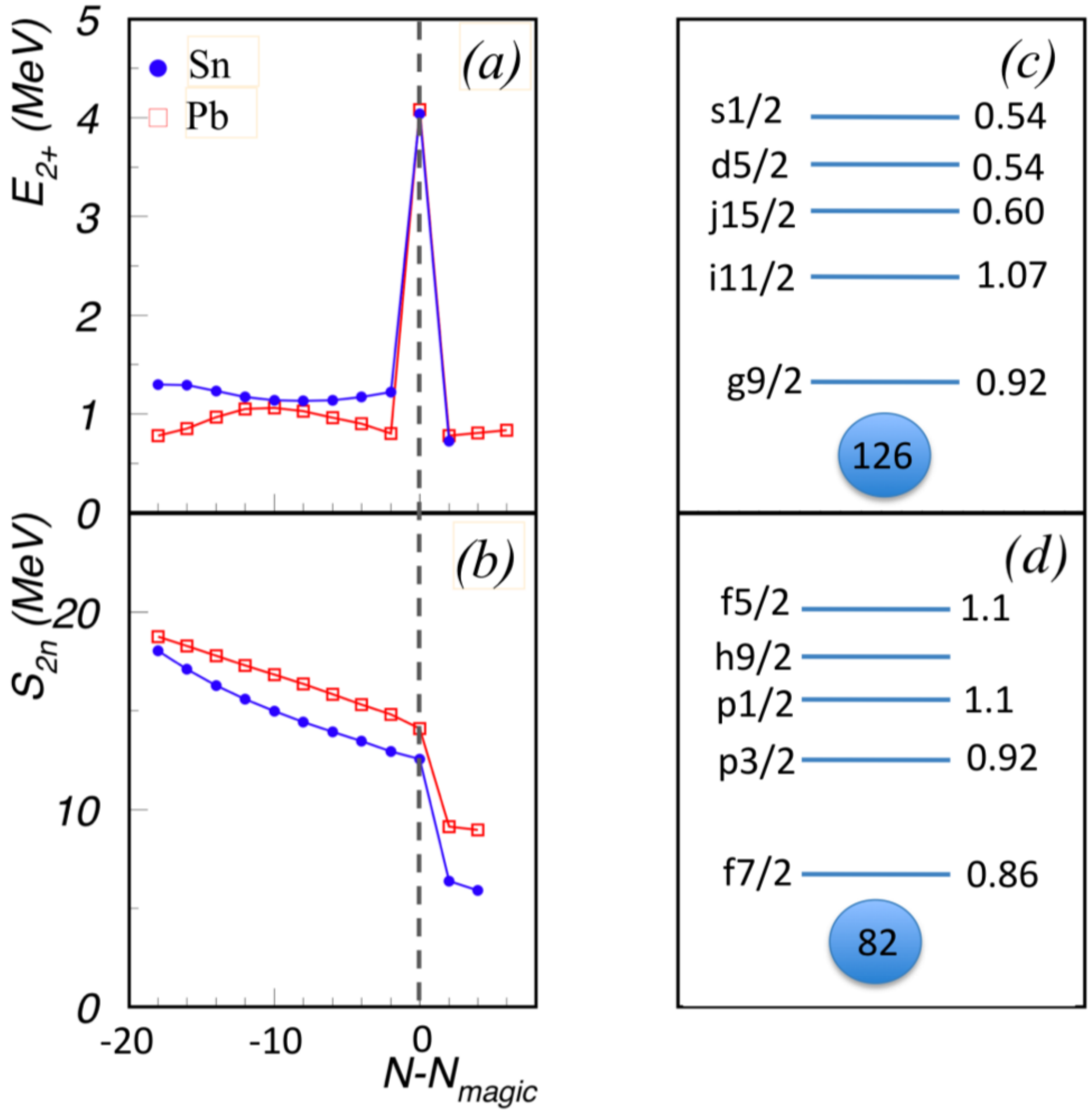

**Figure 1 | Signs of magic nature: comparisons of Pb and Sn isotopes. a**, **b**, Shell closures reveal themselves as discontinuities in the energy of the first electric quadrupole $2^+$ state (**a**) and $S_{2n}$ (**b**). In both cases the observable is plotted against the number of neutrons beyond the shell closure. The best indicator of magic nature lies in the single-particle states outside the closed shell. **c**, **d**, The single-particle states above the magic numbers $N = 126$ (that is, above $^{208}Pb$) (**c**) and $N = 82$ (that is, above $^{132}Sn$) (**d**) are shown, together with the spectroscopic factors. The data were taken from the ENSDF database (http://www.nndc.bnl.gov/ensdf/), from ref. 23 and from the present work.

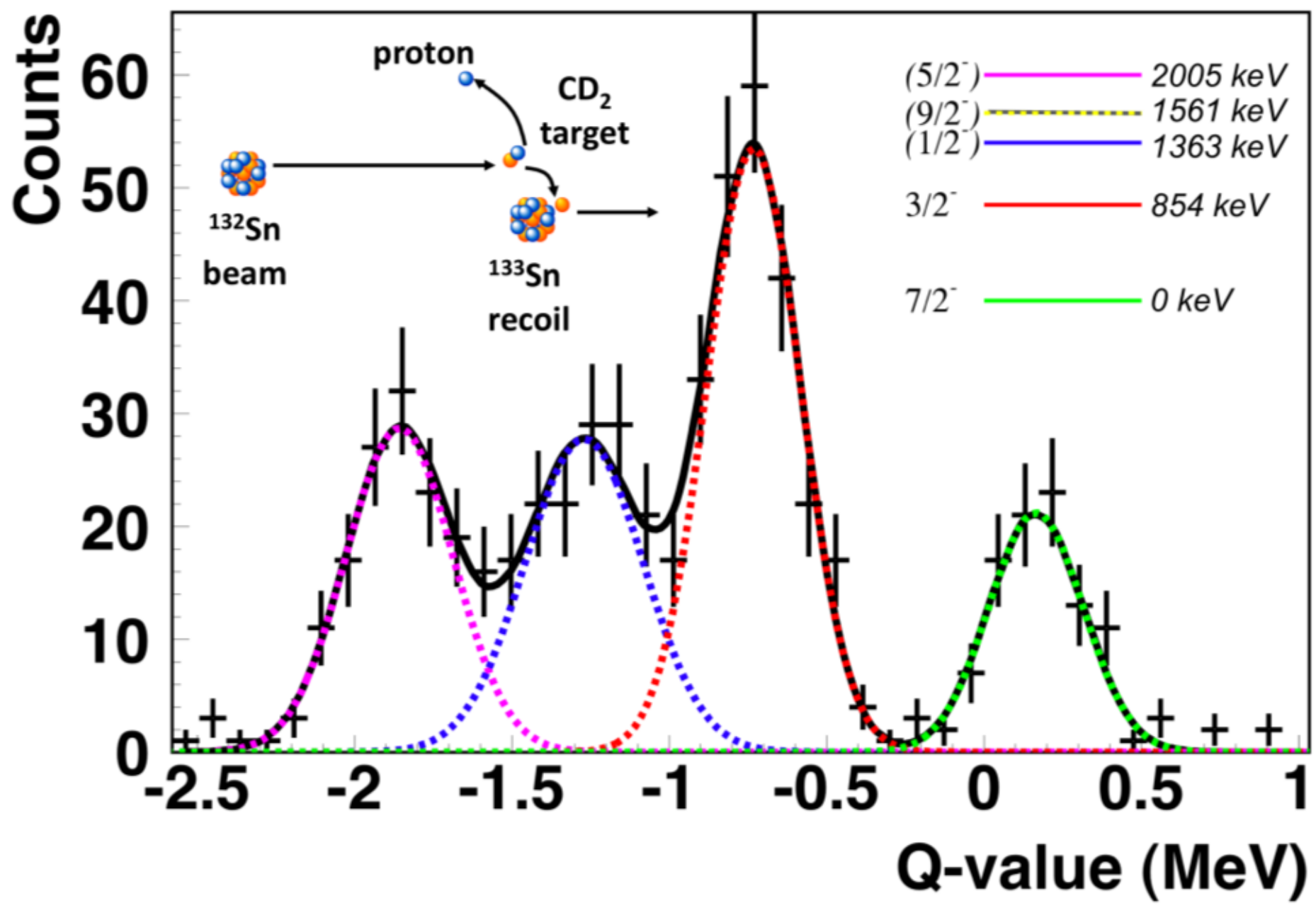

**Figure 2 | *Q*-value spectrum for the $^{132}$Sn(d,p)$^{133}$Sn reaction at 54° in the centre of mass.** Error bars are statistical, shown as a standard deviation in the number of counts. The black solid line shows a fit to four peaks: the ground state (green), the 854-keV state (red), the first observation of the 1363-keV state (blue), and the 2005-keV state (magenta). The top left inset displays a diagram of the (d,p) reaction in inverse kinematics. The top right inset shows the level scheme of $^{133}$Sn. The 1561-keV state, expected to be the $9/2^-$ $h_{9/2}$ state, was not significantly populated in this reaction and therefore was not included in the fit.

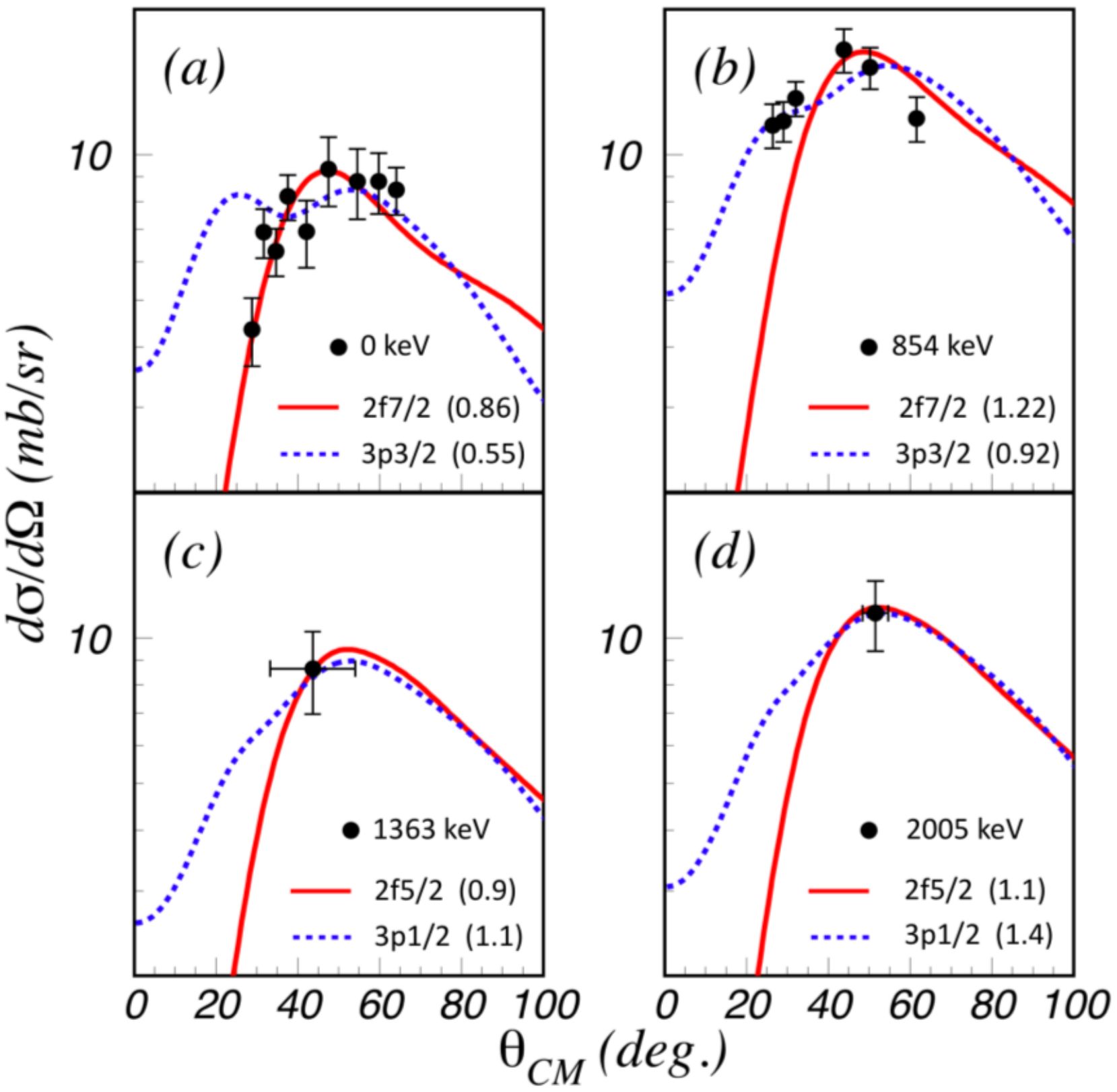

**Figure 3 | Angular distributions, expressed as differential cross sections ($d\sigma/d\Omega$), of protons in the centre of mass resulting from the $^{132}Sn(d,p)^{133}Sn$ reaction for the two lowest states populated and cross-section measurements, also expressed as differential cross sections, for the two highest states.** Calculations for the nearest expected *f*-wave and *p*-wave single-neutron states are shown in red (solid) and blue (dotted), respectively. Error bars refer to the standard deviations in the differential cross-sections. The numbers in parentheses give the spectroscopic factors used to fit the calculation to the data. **a**, Ground state; **b**, 854-keV state; **c**, 1363-keV state; **d**, 2005-keV state.

**Table 1 |** Properties of the four single-particle states populated by the $^{132}Sn(d,p)^{133}Sn$ reaction

| $E_x$ (keV) | $J^{\pi}$ | Configuration | $S$ | $C^2$ (fm$^{-1}$) |
|---|---|---|---|---|
| 0 | $7/2^-$ | $^{132}Sn_{gs} \otimes \nu_{f7/2}$ | $0.86 \pm 0.16$ | $0.64 \pm 0.10$ |
| 854 | $3/2^-$ | $^{132}Sn_{gs} \otimes \nu_{p3/2}$ | $0.92 \pm 0.18$ | $5.61 \pm 0.86$ |
| $1363 \pm 31$ | $(1/2^-)$ | $^{132}Sn_{gs} \otimes \nu_{p1/2}$ | $1.1 \pm 0.3$ | $2.63 \pm 0.43$ |
| 2005 | $(5/2^-)$ | $^{132}Sn_{gs} \otimes \nu_{f5/2}$ | $1.1 \pm 0.2$ | $(9 \pm 2) \times 10^{-4}$ |

The spectroscopic factors ($S$) were extracted from the data by using the Strömich optical potentials, a radius parameter $r$ = 1.25 and diffuseness $a$ = 0.65 for the neutron bound state wave function. The asymptotic normalization coefficient (ANC) is quoted as $C^2$. All errors are expressed as standard deviations. Excitation energies were taken from the ENSDF database (http://www.nndc.bnl.gov/ensdf/) and the present work.

# Supplemental Information

In order to be able to quantify the sensitivity to the optical potentials, the DWBA calculations were repeated for the $^{132}Sn(d,p)^{133}Sn$ reaction using global optical potentials: Lohr-Haeberli[25] for the deuteron incoming wave, and Chapel-Hill 89[26] for the proton outgoing wave. The resulting angular distributions were found to have very similar shapes to those using local optical potentials and agree in magnitude to better than 15% (see figure 4).

For comparison, we also performed calculations for $^{209}Pb$. Data for $^{208}Pb(d,p)^{209}Pb$ at energies close to the Coulomb barrier[23] were used to extract the associated spectroscopic factors. Local and global optical potentials from the same references as for the $^{132}Sn(d,p)^{133}Sn$ were used. The calculations using the Strömich (local) potentials[20] agree well with the shape of the data for the three states in $^{209}Pb$ where data were available (see figure 5). In contrast, calculations with global potentials[25,26] are in poor agreement with these data, especially for large angular momentum transfers. The values of S extracted here for $^{209}Pb$ agree within uncertainties with those from the 1998 analysis of $^{208}Pb(d,p)$ data by Hirota et al.[24]. The spectroscopic factors for $^{209}Pb$ in figure 1 come from the current reanalysis of the Ellegaard et al[23] data for the ground state, 779 keV and 1423 keV excited states, using the Strömich potentials. The S values for the 1557 keV and 2032 keV states come from Hirota et al[24]. It was not possible to reanalyze these states using the Strömich potentials, as data in the beam energy range required for this choice of potentials were not available.

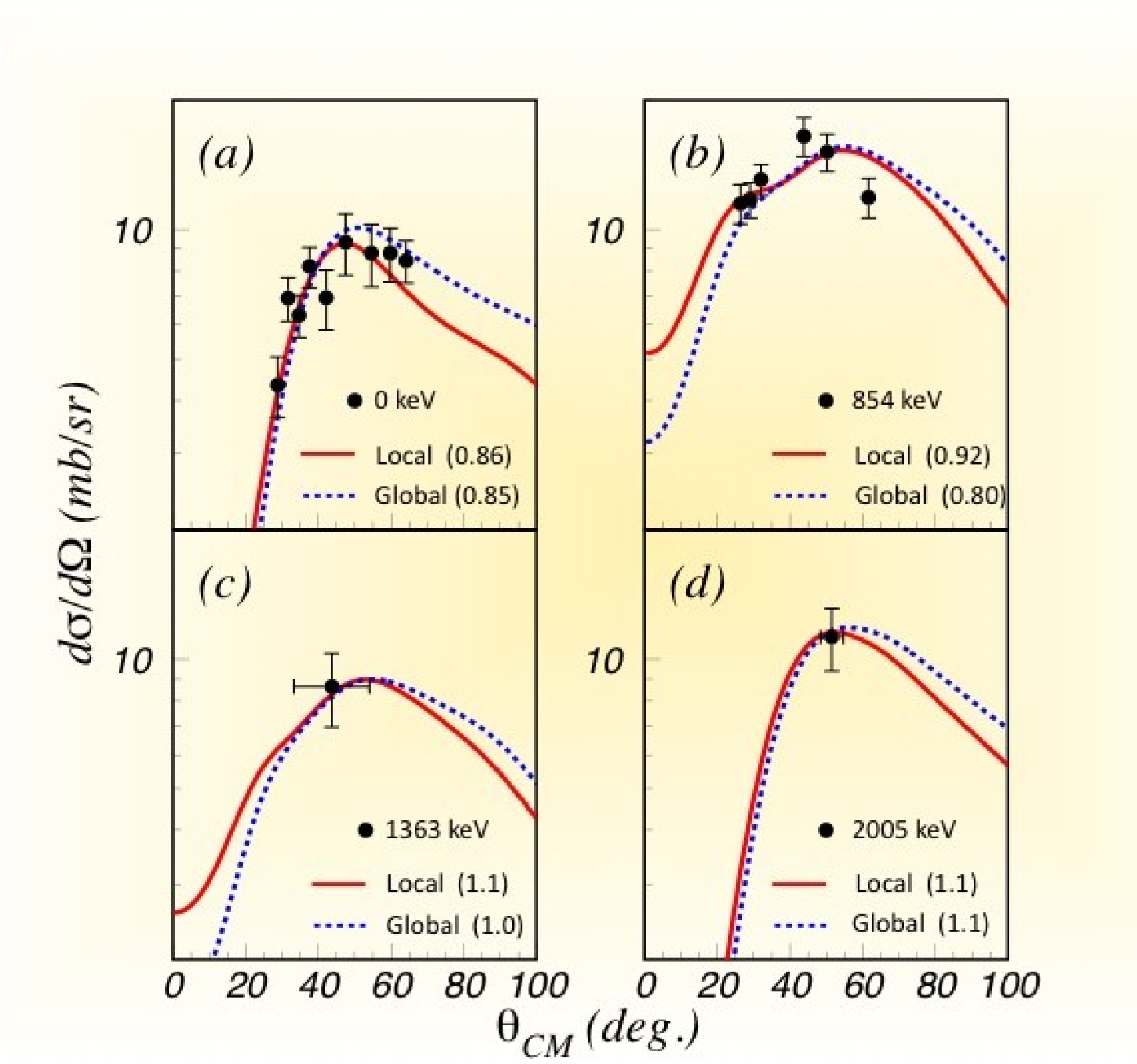

**Figure 4 Comparison of calculations of the $^{132}Sn(d,p)^{133}Sn$ reaction using Local (Strömich) and Global (Lohr-Haeberli + Chapel-Hill 89) optical potentials.** The numbers in parentheses are the spectroscopic factors extracted. The four frames correspond to the four individual states: **a** ground state, **b** 854 keV state**, c** 1363 keV state and **d** 2005 keV state.

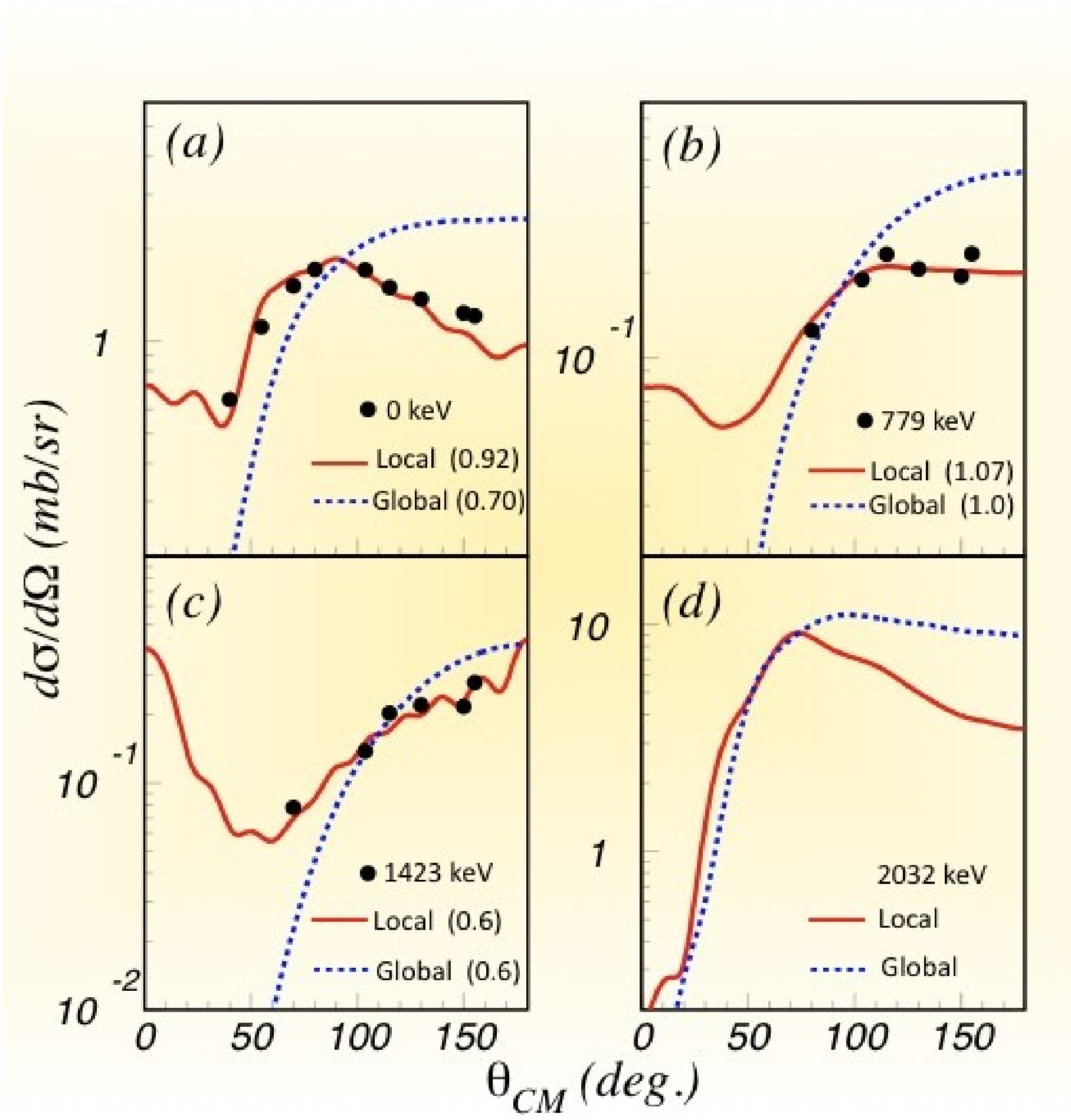

**Figure 5 Comparison of calculations of the $^{208}$Pb(d,p)$^{209}$Pb reaction using Local (Strömich) and Global (Lohr-Haeberli + Chapel-Hill 89) optical potentials.** The numbers in parentheses are the spectroscopic factors extracted. The four frames correspond to the four individual states: **a** ground state (9/2+), **b** 779 keV state (11/2+)**, c** 1423 keV state (15/2-)and **d** 2032 keV (1/2+) state. Data[23] for deuterons at energies close to the Coulomb barrier are available only for the lowest three states.

## The data shown in figure 3 are tabulated below.

Table 2 Differential cross sections for population of the ground state of $^{133}$Sn in the $^{132}$Sn(d,p)$^{133}$Sn reaction. The data are plotted in figure 3 panel a. Statistical uncertainties quoted represent the standard deviation in the differential cross section.

| $\theta_{CM}$ (deg.) | dσ/dΩ (mb/sr) |
|---|---|
| 28.8 | 4.3 ± 0.7 |
| 31.7 | 6.9 ± 0.8 |
| 34.7 | 6.3 ± 0.7 |
| 37.6 | 8.2 ± 0.9 |
| 42.1 | 6.9 ± 1.1 |
| 47.5 | 9.3 ± 1.5 |
| 54.5 | 8.8 ± 1.5 |
| 59.8 | 8.8 ± 1.3 |
| 64.1 | 8.4 ± 0.9 |

Table 3 Differential cross sections for population of the 854 keV excited state of $^{133}$Sn in the $^{132}$Sn(d,p)$^{133}$Sn reaction. The data are plotted in figure 3 panel b. Statistical uncertainties quoted represent the standard deviation in the differential cross section.

| $\theta_{CM}$ (deg.) | dσ/dΩ (mb/sr) |
|---|---|
| 26.4 | 11.5 ± 1.2 |
| 29.1 | 11.7 ± 1.1 |
| 32.1 | 13.1 ± 1.1 |
| 43.8 | 16.5 ± 1.7 |
| 50.2 | 15.1 ± 1.5 |
| 61.6 | 11.9 ± 1.3 |

Table 4 Cross sections for population of the 1353 keV and 2005 keV excited states of $^{133}$Sn in the $^{132}$Sn(d,p)$^{133}$Sn reaction at the angular range given. The data are plotted in figure 3 panels c and d. Statistical uncertainties quoted represent the standard deviation in the angle integrated cross section.

| $E_x$ (keV) | Range of $\theta_{CM}$ (deg.) | $d\sigma/d\Omega$ (mb/sr) |
|---|---|---|
| 1353 | 33.3 to 54.0 | 8.7 ± 1.7 |
| 2005 | 48.4 to 54.7 | 11.3 ± 1.9 |

25. Lohr, J.M. & Haeberli, W, Elastic scattering of 9-13 MeV vector polarized deuterons, *Nucl. Phys.* **A 232**, 381 - 397 (1974).

26. Varner, R., Thompson, W.J., McAbee, T.L., Ludwig, E.J., & Clegg, T.B., A global nucleon optical model potential, *Phys. Rep.* **201**, 57 - 119 (1991).